\documentclass[prc,twocolumn,showpacs,amsmath,amssymb,superscriptaddress,floatfix,nofootinbib,10pt]{revtex4}
\usepackage{mathrsfs,bm}
\usepackage{longtable,lscape}
\usepackage{txfonts}
\usepackage{amssymb}
\usepackage{indentfirst}
\usepackage{graphicx,,booktabs}
\usepackage{multirow}
\usepackage{overpic}
\usepackage{color}
\usepackage{amssymb}

\begin{document}
\title{Magnetic moments of the spin-1/2 doubly charmed baryons in covariant baryon chiral perturbation theory}

\author{Ming-Zhu Liu}
\affiliation{School of Physics and
Nuclear Energy Engineering \& International Research Center for Nuclei and Particles in the Cosmos \&
Beijing Key Laboratory of Advanced Nuclear Materials and Physics,  Beihang University, Beijing 100191, China}

\author{Yang Xiao}
\affiliation{School of Physics and
Nuclear Energy Engineering \& International Research Center for Nuclei and Particles in the Cosmos \&
Beijing Key Laboratory of Advanced Nuclear Materials and Physics,  Beihang University, Beijing 100191, China}

\author{Li-Sheng Geng}
\email[E-mail: ]{lisheng.geng@buaa.edu.cn} \affiliation{School of
Physics and Nuclear Energy Engineering \& International Research
Center for Nuclei and Particles in the Cosmos \& Beijing Key
Laboratory of Advanced Nuclear Materials and Physics,  Beihang
University, Beijing 100191, China}
\begin{abstract}
Inspired by the recent discovery of the $\Xi_{cc}^{++}$ by the LHCb
Collaboration, we study the magnetic moments of the spin-1/2
doubly charmed baryons up to the next-to-leading order in
covariant baryon chiral perturbation theory with the
extended-on-mass-shell renormalization scheme. There are three
low energy constants at this order, $a_1$, $a_2$ and $g_a$. The latest lattice QCD simulations allow us to fix a combination of $a_1$ and $a_2$, while
the axial-vector coupling $g_a$ can be determined  in three different ways, either by fitting to the lattice QCD data, or by the quark model, or by the heavy antiquark diquark symmetry. The
magnetic moments of the spin-1/2 doubly charmed baryons $\Xi^d_{cc}$ and $\Xi^s_{cc}$ can then
be predicted. We compare our results with those obtained in the
heavy baryon chiral perturbation theory and other approaches,  and point out some inconsistencies between the lattice QCD simulations and the quark model.

\end{abstract}


\maketitle
\section{INTRODUCTION}

The doubly charmed baryons, $\Xi_{cc}^{u}$, $\Xi_{cc}^{d}$ and
$\Xi_{cc}^{s}$, are composed of two charm quarks and one light quark. One of them, $\Xi_{cc}^{+}$, with a mass of $3519\pm2$ MeV was first reported by the SELEX Collaboration~\cite{Mattson:2002vu,Ocherashvili:2004hi}. Unfortunately,
no other collaborations found such  a state. Recently, the LHCb Collaboration observed another doubly charmed baryon state $\Xi_{cc}^{++}$ with a mass of $3621.4\pm0.78$ MeV, which has inspired many  theoretical studies on its  weak~\cite{Yu:2017zst,Wang:2017mqp,Wang:2017azm}, strong and radiative decays~\cite{Li:2017pxa,Xiao:2017udy,Cui:2017udv}.

The magnetic moment of a hadron is one of its most important properties, which encodes crucial information on its inner structure.  In the past, many phenomenological  models have been used to study the magnetic moments of $\Xi_{cc}$~\cite{Lichtenberg:1976fi,JuliaDiaz:2004vh,Oh:1991ws,Patel:2008xs,Sharma:2010vv,Faessler:2006ft,SilvestreBrac:1996bg,Bose:1980vy,Jena:1986xs}. More recently, they have been calculated in
heavy baryon chiral perturbation theory (HB ChPT)~\cite{Li:2017vmq} and QCD sum rules~\cite{Ozdem:2018uue}.   In this work, we will study the magnetic moments of the spin-1/2
doubly charmed baryons up to the next-to-leading  order (NLO) in
covariant baryon chiral perturbation theory (BChPT) with the
extended-on-mass-shell (EOMS) renormalization scheme.  In the present work, we will contrast the ChPT results with the lattice QCD data of Ref.~\cite{Can:2013tna} to determine the unknown low energy constants (LECs). In many recent studies (see, e.g., Refs.~\cite{Ren:2012aj,Xiao:2018rvd}), it has been shown that the EOMS  BChPT can provide a better description of the lattice QCD quark-mass dependent results than its non-relativistic
 counterpart.

Chiral perturbation theory (ChPT) is a low energy effective field theory of  QCD, which plays an important role in our understanding of the non-perturbative strong interaction.
In ChPT, relevant Feynman diagrams contributing to a certain process are organized as an expansion in powers of the external momenta and  light quark masses. In the center of such an expansion is a power counting scheme, first proposed by Weinberg~\cite{Weinberg:1978kz}. However, in the one-baryon sector,
the naive power counting breaks down because of the large non-zero baryon
mass  $m_0$ in the chiral limit.
To overcome this issue, HB ChPT was  proposed~\cite{Jenkins:1990jv,Bernard:1995dp}, which performs a dual expansion
  in terms of both $1/m_0$  and the chiral expansion. Later, two relativistic schemes were also proposed, i.e., the infrared (IR)~\cite{Becher:1999he} and EOMS~\cite{Fuchs:2003qc} schemes.
For a recent and concise summary of different schemes, see, e.g., Ref.~\cite{Geng:2013xn}.

The EOMS scheme has already been successfully applied to study many physical observables such as the magnetic moments~\cite{Geng:2009ys,Geng:2008mf,Geng:2009hh,Xiao:2018rvd},
the masses and sigma terms~\cite{Ren:2012aj,Ren:2014vea,Ling:2017jyz,Ren:2013oaa} of the octet and decuplet baryons, the hyperon vector couplings~\cite{Geng:2009ik,Geng:2014efa},
the axial vector charges~\cite{Ledwig:2014rfa}, the pion-nucleon scattering~\cite{Alarcon:2012kn,Chen:2012nx},  the nucleon Compton scattering~\cite{Lensky:2009uv},  the neutral pion photo production~\cite{Blin:2014rpa},
the scattering of pseudoscalar mesons off $D/B$ mesons~\cite{Geng:2010vw,Altenbuchinger:2013vwa,Yao:2015qia}, the $DD^*$ scattering~\cite{Xu:2017tsr}, and the $\Xi_{cc}$ masses and sigma terms~\cite{Sun:2016wzh,Yao:2018ifh}.
It will be interesting to see how it describes the magnetic moments of the $\Xi_{cc}$ baryons particularly from the perspective of lattice QCD simulations.

This work is organized as follows. In Section II, we provide the theoretical ingredients and calculate the pertinent Feynman diagrams. Results and discussions are given in
Section III, followed by a short summary in Section IV.

\section{Theoretical formalism}
\begin{figure}[!h]
\centering
\begin{overpic}[scale=.51]{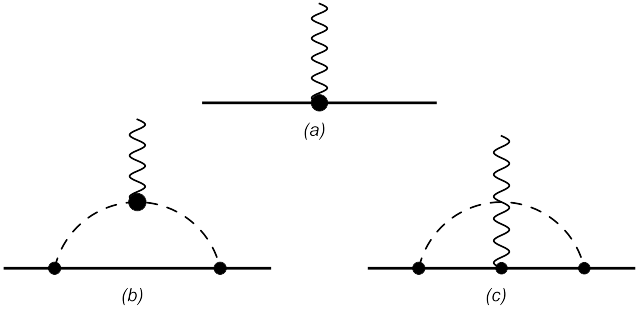}
\end{overpic}
\caption{Feynman diagrams contributing to the magnetic moments of the $\Xi_{cc}$ baryons: (a) tree level, (b) meson pole  and (c) baryon pole.  The solid lines denote the
doubly charmed baryons, the dashed lines denote the Nambu-Goldstone bosons, and the wiggly lines indicate the photon. The heavy dots indicate $\mathcal{O}(p^2)$ vertices and
the normal dots denote $\mathcal{O}(p)$ vertices.}\label{f2}
\end{figure}

The magnetic moments of doubly charmed baryons are defined via
the  matrix elements of the electromagnetic current $J_{\mu}$ in the following way,
\begin{eqnarray}
 \nonumber
\langle \Psi(p^{\prime})|J_{\mu}|\Psi(p)\rangle &=&
\bar{u}(p^{\prime})[\gamma_{\mu}F_{1}^{B}(t)+\frac{i\sigma_{\mu\nu}q^{\nu}}{2
m_{B}}F_{2}^{B}(t)]u(p), \label{def1}
\end{eqnarray}
where $\bar{u}(p^{\prime})$ and $u(p)$ are Dirac spinors, $m_B$ is
the chiral limit doubly charmed baryon mass,  and $F_{1}^{B}(t)$
and $F_{2}^{B}(t)$ denote the Dirac and Pauli form factors,
respectively. The four-momentum transfer is defined as
$q=p'-p$ and $t=q^2$. At $t=0$, $F_{2}^{B}(0)$ is the
so-called anomalous magnetic moment, $\kappa_{B}$, and the
magnetic moment is $\mu_{B}=\kappa_{B}+Q_{B}$, with $Q_{B}$ the
charge of the doubly charmed baryon. Up to NLO, there are three Feynman diagrams contributing to the magnetic moments of the $\Xi_{cc}$ as shown in
Fig.~\ref{f2}, where Diagram (a) is of $\mathcal{O}(p^2)$ and Diagrams (b) and (c) are of $\mathcal{O}(p^3)$.

\subsection{Tree level diagram}

The leading order (tree-level) contribution is provided by the following
Lagrangian,
\begin{eqnarray}
\mathcal{L}^{(2)}_{MB}&=&a_{1}\frac{1}{8m_{B}}\bar{H}\sigma^{\mu\nu}\hat{F}_{\mu\nu}^{+}H
+a_{2}\frac{1}{8m_{B}}\bar{H}\sigma^{\mu\nu}H
Tr({F}_{\mu\nu}^{+}),
\end{eqnarray}
where $\sigma^{\mu\nu}=\frac{i}{2}[\gamma^{\mu},\gamma^{\nu}]$,
$F_{\mu\nu}^{+}=|e|(u^{\dag}Q' F_{\mu\nu}u+uQ'F_{\mu\nu}u^{\dag})$,
$F_{\mu\nu}=\partial_{\mu}A_{\nu}-\partial_{\nu}A_{\mu}$,
$\hat{F}_{\mu\nu}^{+}=F_{\mu\nu}^{+}-\frac{1}{3}Tr(F_{\mu\nu}^{+})$,
and $Q'=$diag(2,1,1) is the baryon charge matrix,
$u=\exp[i\Phi/2f_{\phi}]$ with $\Phi$ the unimodular matrix
containing the pseudoscalar nonet and $f_{\phi}$ the pseudoscalar
decay constant. In the numerical analysis, we use the following physical values for the decay constants: $f_\pi=92.4$ MeV, $f_K=1.22 f_\pi$, $f_\eta=1.3 f_\pi$. For $m_B$, we use the SU(3) average of the lattice QCD results,
i.e. $m_B= 3722$ MeV~\cite{Can:2013tna}. The $\Xi_{cc}$ baryons are contained in a column $H$, which reads
\begin{eqnarray}
H=\left(\begin{matrix}
&\Xi_{cc}^{u} \\
&\Xi_{cc}^{d}  \\
&\Xi_{cc}^{s}
\end{matrix}\right).
\end{eqnarray}
The tree level contributions to the magnetic moments can be easily obtained as
\begin{eqnarray}\label{tree}
\mu_{B}^{(2)}=\alpha_{B}a_{1}+\beta_{B}a_{2},
\end{eqnarray}
where
$\alpha_{B}=(\langle\bar{H}Q'H\rangle-\frac{1}{3}\bar{H}H\langle
Q'\rangle)$ and  $\beta_{B}=\bar{H}H\langle Q'\rangle$ are
given in Table.~\ref{t1}. We will determine the two LECs $a_{1}$ and $a_{2}$ by fitting 
to the lattice QCD simulations.

 \begin{table}[!h]
 \centering
  \caption{$\mathcal{O}(p^2)$ coefficients appearing in Eq.~(\ref{tree}).}\label{t1}
 \begin{tabular}{c|ccc}
 \hline\hline
  &$\Xi_{cc}^{u}$  & $\Xi_{cc}^{d}$ &  $\Xi_{cc}^{s}$   \\
\hline
$\alpha_{B}$           & 2/3       &$-1/3$      &$-1/3$                                                 \\
$\beta_{B}$           &4      &4      &4                                               \\
           \hline\hline
 \end{tabular}
 \end{table}

\subsection{Loop diagrams}

At $\mathcal{O}(p^3)$, there are two Feynman diagrams, the so-called baryon-pole and meson-pole  diagrams, as shown in
Fig.~\ref{f2}.

The Lagrangian for a  doubly charmed baryon interacting with a Nambu-Goldstone boson (NGB)  is
\begin{eqnarray}
\mathcal{L}^{(1)}_{MBB}&=&\frac{g_{a}}{2}\bar{H}\gamma^{\mu}\gamma^{5}u_{\mu}H,
\end{eqnarray}
where
$u_{\mu}=[u^{\dag}(\partial_{\mu}-ir_{\mu})u-u(\partial_{\mu}-il_{\mu})u^{\dag}]$, and $g_a$ is the axial-vector coupling constant.

The Lagrangian describing the interaction between a baryon and a photon is  of $\mathcal{O}(p)$ and reads
\begin{eqnarray}
\mathcal{L}_{B}^{(1)}=i\bar{H}\gamma^{u}D_{\mu}H,
\end{eqnarray}
where $D_{\mu}=\partial_{\mu}
+\Gamma_{\mu}$, $\Gamma_{\mu}=\frac{1}{2}[u^{\dag}(\partial_{\mu}-ir_{\mu})u+u(\partial_{\mu}-il_{\mu})u^{\dag}]
=\frac{1}{2}(u^{\dag}\partial_{\mu}u+u\partial_{\mu}u^{\dag})-\frac{i}{2}(u^{\dag}r_{\mu}u+ul_{\mu}u^{\dag})=-ieQ'A_{\mu}$.

The Lagrangian describing the interaction between a meson and a photon is of $\mathcal{O}(p^2)$ and reads,
\begin{eqnarray}
\mathcal{L}^{(2)}_{M}&=&\frac{f_{\phi}^{2}}{4}Tr[\bigtriangledown_{\mu}U(\bigtriangledown^{\mu}U)^{\dag}]
\end{eqnarray}
where
$\bigtriangledown_{\mu}U=\partial_{\mu}U+ieA_{\mu}(QU-UQ)$ and
$Q=\mathrm{diag}(2/3,-1/3,-1/3)$.

From these, one can easily obtain the loop contributions
to the magnetic moments, i.e.,
\begin{equation}\label{loop}
\mu^i_\mathrm{loop}=c_b^i(\phi) H^b(m_\phi)+ c_m^i(\phi) H^m(m_\phi),
\end{equation}
where $c_b^i(\phi)$ and $c_m^i(\phi)$ are tabulated in Tables~\ref{t2} and \ref{t3}, $i$ runs over $\Xi_{cc}^{u}$,
$\Xi_{cc}^{d}$, and $\Xi_{cc}^{s}$, and $\phi$ denotes $\pi$, $K$
or $\eta$. The loop functions $H^b(m_\phi)$ and $H^m(m_\phi)$ with
$m_\phi$ the mass of a NGB are:
\begin{eqnarray}
H^{b}(m_\phi)&=&-\frac{g_{a}^2}{16\pi^2f_{\phi}^2}\left[m_{B}^{2}+2m_{\phi}^{2}+\frac{m_{\phi}^2}{m_{B}^2}(m_{B}^2-m_{\phi}^2)\log\left(\frac{m_{\phi}^2}{m_{B}^2}\right)
  \right.        \nonumber \\
&+&\left.\frac{2m_{\phi}^3(m_{\phi}^2-3m_{B}^2)}{m_{B}^{2}\sqrt{4m_{B}^2-m_{\phi}^2}}\arccos(\frac{m_{\phi}}{2m_{B}})\right],
\end{eqnarray}
\begin{eqnarray}
H^{m}(m_\phi)&=&\frac{g_{a}^2}{16\pi^2f_{\phi}^2}\left[-m_{B}^2+2m_{\phi}^2+\frac{m_{\phi}^2}{m_{B}^2}(2m_{B}^2-m_{\phi}^2)\log\left(\frac{m_{\phi}^2}{m_{B}^2}\right)\right.
\nonumber \\
&+&\left.\frac{2m_{\phi}(m_{\phi}^4-4m_{\phi}^2m_{B}^2+2m_{B}^2)}{m_{B}^2\sqrt{4m_{B}^2-m_{\phi}^2}}
\arccos(\frac{m_{\phi}}{2m_{B}})\right].
\end{eqnarray}

Up to NLO,  the total magnetic moments are a sum of the tree and loop contributions and
they are usually expressed in units of the nucleon magneton $\mu_N$. In the end we obtain
\begin{eqnarray*}\label{mm_total}
\mu_{\Xi_{cc}^{u}}&=&\frac{m_{N}}{m_B}(2+\frac{2}{3}a_{1}+4a_{2}+c^1_{b}(\phi)H^{b}(m_\phi)+c^1_{m}(\phi)H^{m}(m_\phi)),
\\ \nonumber
\mu_{\Xi_{cc}^{d}}&=&\frac{m_{N}}{m_B}(1-\frac{1}{3}a_{1}+4a_{2}+c^2_{b}(\phi)H^{b}(m_\phi)+c^2_{m}(\phi)H^{m}(m_\phi)),
\\ \nonumber
\mu_{\Xi_{cc}^{s}}&=&\frac{m_{N}}{m_B}(1-\frac{1}{3}a_{1}+4a_{2}+c^3_{b}(\phi)H^{b}(m_\phi)+c^3_{m}(\phi)H^{m}(m_\phi)),
\\ \nonumber
\end{eqnarray*}
where $m_N=940$ MeV is the nucleon mass.

 \begin{table}[!h]
 \centering
  \caption{Coefficients of the baryon-pole contributions appearing in Eq.~(\ref{loop}). }\label{t2}
 \begin{tabular}{c|ccc}
 \hline\hline
$c_{b}$  &$\Xi_{cc}^{u}$  & $\Xi_{cc}^{d}$ &  $\Xi_{cc}^{s}$   \\ \hline
$\pi$           & 4       &5     &0                                                 \\
$\eta$           &2/3       &1/3     &4/3                                               \\
$K$           & 2       &$2$      &6                                                \\
           \hline\hline
 \end{tabular}
 \end{table}

\begin{table}[!h]
 \centering
  \caption{Coefficients of the meson-pole contributions appearing in Eq.~(\ref{loop}). }\label{t3}
 \begin{tabular}{c|ccc}
 \hline\hline
$c_{m}$  &$\Xi_{cc}^{u}$  & $\Xi_{cc}^{d}$ &  $\Xi_{cc}^{s}$   \\ \hline
$\pi$           & $-2$       &2      &0                                                 \\
$K$           &$-2$      &0      &2                                                \\
           \hline\hline
 \end{tabular}

 \end{table}

\section{ Results and discussions}

 In the following, we determine the LECs $a_1$ and $a_2$ by fitting to the lattice QCD simulations of Ref.~\cite{Can:2013tna}, which are given in Table~\ref{Lattice}.
 The LEC $g_a$ will be determined by three ways, either (Case 1) by fitting to the lattice QCD simulations , (Case 2)  by the heavy antiquark diquark symmetry (HADS)  or (Case 3 by the quark model ).
 To quantify the agreement with the lattice QCD data,  we use  the $\chi^2$ defined as
 \begin{equation}
 \chi^2_j=\sum_{k=1}^4 \frac{(\mu^k_{theo.}-\mu^k_{lQCD})^2}{d_k^2},
 \end{equation}
 where $\mu^k_{theo.}$ and $\mu^k_{lQCD} (d_k)$ are the magnetic moments (uncertainties) obtained in BChPT and those of the lattice QCD simulations of Table \ref{Lattice} for
 $\Xi_{cc}^d$ ($j=1$) and $\Xi_{cc}^s$ ($j=2$) , respectively.

From Eq.~(\ref{mm_total}), it is clear that since the lattice QCD data are only available for $\Xi_{cc}^d$ and $\Xi_{cc}^s$, we cannot determine the LECs $a_1$ and $a_2$ simultaneously. Only the combination
$c_1=-\frac{1}{3}a_1+4a_2$ can be fixed. As a result, we cannot predict the magnetic moment of $\Xi^u_{cc}$ without further inputs.

\begin{table}[!h]
 \centering
  \caption{  Lattice QCD magnetic moments and masses of $\Xi_{cc}^{d}$ and $\Xi_{cc}^{s}$  at different $m_{\pi}^2$~\cite{Can:2013tna}.
 }\label{Lattice}
 \begin{tabular}{c|ccccc}
 \hline\hline
  & $m_{\pi}^2$     &$m_{\Xi_{cc}^{d}}$     &$m_{\Xi_{cc}^{s}}$           &$\mu_{\Xi_{cc}^{d}}$    &
$\mu_{\Xi_{cc}^{s}}$   \\
\hline
\multirow{4}{0.8cm}{Latt}&0.490  &3.810(12)  &3.861(17)   &0.412(13)     &0.389(18) \\
                         &0.325   &3.740(13)  &3.806(12)   &0.404(12)     &0.386(11) \\
                         &0.168   &3.708(16)  &3.788(16)   &0.410(20)     &0.400(11)  \\
                         &0.090   &3.689(18)  &3.781(28)   &0.416(19)     &0.402(15)  \\
           \hline\hline
 \end{tabular}
 \end{table}

\subsection{ Results at $\mathcal{O}(p^2)$}

If we just consider the tree level contribution,  we have only one LEC, $c_{1}$. It can be determined by fitting
to the lattice QCD data. The resulting  value  and $\chi^2$ are shown in Table~\ref{treefit}. The predicted magnetic moments of $\Xi_{cc}$ at the physical pion mass  are
\begin{equation}
\mu_{\Xi_{cc}^{d}}=\mu_{\Xi_{cc}^s}=\frac{m_{N}}{m_B}(c_1+1)=0.401(3)\,\mu_{N},
\end{equation}
where the number in the parenthesis is the uncertainty at the 68\% confidence level.

 \begin{table}[!h]
 \centering
  \caption{ $\mathcal{O}(p^2)$ LEC determined by  fitting to the lattice QCD data of Table \ref{Lattice}~\cite{Can:2013tna} and the corresponding $\chi^2$.}\label{treefit}
 \begin{tabular}{c|ccc}
 \hline\hline
  $\mathcal{O}(p^2)$&$c_{1}$    & $\chi_{\Xi_{cc}^{d}}$ &  $\chi_{\Xi_{cc}^{s}}$  \\ \hline
          & $0.586(19)$               &1.678  &2.238                                                \\
           \hline\hline
 \end{tabular}
 \end{table}

\subsection{Results at $\mathcal{O}(p^3)$}

At $\mathcal{O}(p^3)$, the meson masses will contribute via the loop diagrams. We determine the eta and kaon masses
by leading order ChPT. Setting the strange quark mass at its physical value, we obtain the following relation:
\begin{eqnarray}
m_{K}^2&=&\frac{1}{2}m_{\pi}^2+\left(m_{K}^2-\frac{1}{2}m_{\pi}^2\right)_\mathrm{phys},   \\ \nonumber
m_{\eta}^2&=&\frac{1}{3} m_{\pi}^2+\frac{4}{3}\left(m_{K}^2-\frac{1}{2}m_{\pi}^2\right)_\mathrm{phys}.
\end{eqnarray}

Fitting to the lattice QCD simulations tabulated in Table ~\ref{Lattice} and with the LEC $g_a$ determined in the three different ways explained above, the resulting LECs as well as the  $\chi^{2}$
 are tabulated in Table~\ref{ggdd1}.
For the sake of comparison we show as well the results obtained  in HBChPT. It is seen  that the lattice QCD
data seem to prefer a $g_a$ smaller than that predicted either by the quark model or the HADS. Furthermore,
as $g_a$ becomes larger,  the EOMS BChPT description of the lattice QCD data becomes slightly better than that of the HBChPT, although for Case 1, where $g_a$ is
taken as a free LEC, the descriptions are of similar quality.

  \begin{table}[!h]
 \centering
  \caption{ Low energy constants $c_1$ and $g_a$  and the corresponding $\chi^2$  of each case described in the text. }\label{ggdd1}
 \begin{tabular}{c|cc|cc|cc}
 \hline\hline
\multicolumn{1}{c}{} & \multicolumn{2}{c}{Case 1}  &  \multicolumn{2}{c}{Case 2}  &  \multicolumn{2}{c}{Case 3}\\
\hline
           &{EOMS} & {HB}     &{EOMS} & {HB}  &{EOMS} & {HB} \\
\hline
     $c_{1}$               & {0.535(82)}             & {0.542(70)}      & {0.249 (19)}             & {0.264(19) } & {0.060(19) }        & {0.083(21) } \\
     $g_a$            &{0.078(61) }             &{0.074(56) }  &{0.2 }         &{0.2 } & {0.25 }       & {0.25 } \\
   $\chi_{\Xi_{cc}^{d}}^2$              &1.494              &1.513   &11.175             &13.180 & 27.797           & 32.785 \\
      $\chi_{\Xi_{cc}^{s}}^2$              &2.039              &2.048  &4.268            &4.448 &8.664          & 9.155\\
           \hline\hline
 \end{tabular}
 \end{table}

\begin{figure}[!h]
\centering
\begin{overpic}[scale=.33]{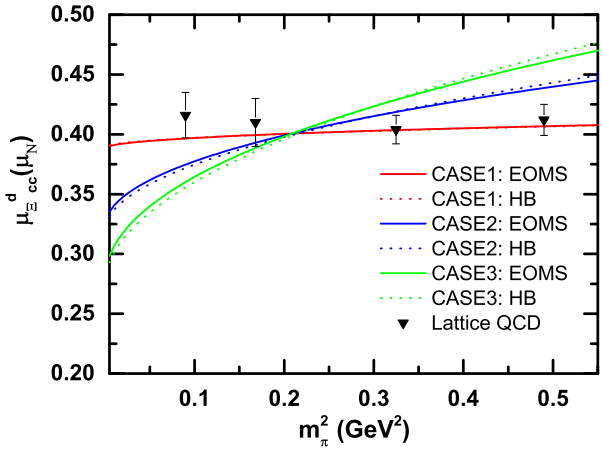}
\end{overpic}
\caption{Magnetic moment of $\Xi_{cc}^{d}$  as a function of $m_{\pi}^2$. The theoretical results are obtained with the LEC $c_1$ determined by
fitting to the lattice QCD data and the LEC $g_a$ determined in three different ways as explained in the text. }\label{d}
\end{figure}

\begin{figure}[!h]
\centering
\begin{overpic}[scale=.33]{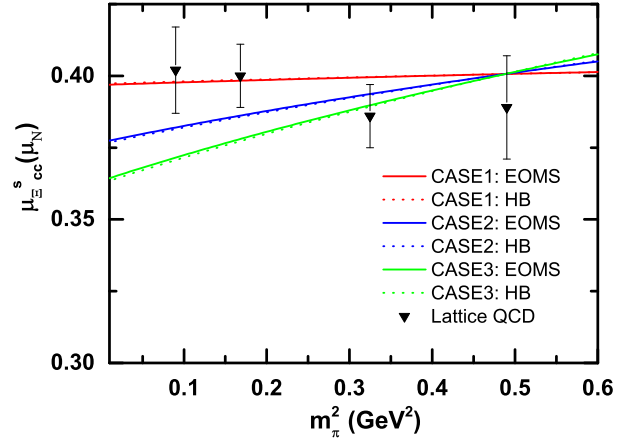}
\end{overpic}
\caption{Same as Fig.~\ref{d}, but for the magnetic moment of $\Xi_{cc}^{s}$. }\label{s}
\end{figure}

In Figs.~\ref{d} and \ref{s}, we plot the predicted magnetic moments of $\Xi_{cc}^{d}$ and $\Xi_{cc}^{s}$  as a function of $m_{\pi}^2$ ,  in comparison with the lattice
QCD data. As can be clearly seen,  there is not much difference between the EOMS and HB results. However, somehow surprisingly, using the $g_a$ determined by either the quark model or the HADS yields unacceptable fits. This indicates that there is considerable discrepancy between the quark model (the antiquark diquark symmetry) and the lattice QCD simulations of Ref.~\cite{Can:2013tna}.~\footnote{One may
need go to the next-to-next-to-leading order (NNLO) to draw a firm conclusion. However, at present this is not feasible because of the increase in the number of free LECs in BChPT and
 the limited lattice QCD data.}

Note that we have used all the 8 sets of lattice QCD data and some of them are obtained with pion masses as large as 700 MeV. They are probably beyond the limit where an $\mathcal{O}(p^3)$ BChPT study can be trusted. Nevertheless, it is clear from the plots that limiting ourselves to the lattice QCD data with smaller pion masses will not change qualitatively any of our conclusions.

\begin{figure}[!h]
\centering
\begin{overpic}[scale=.33]{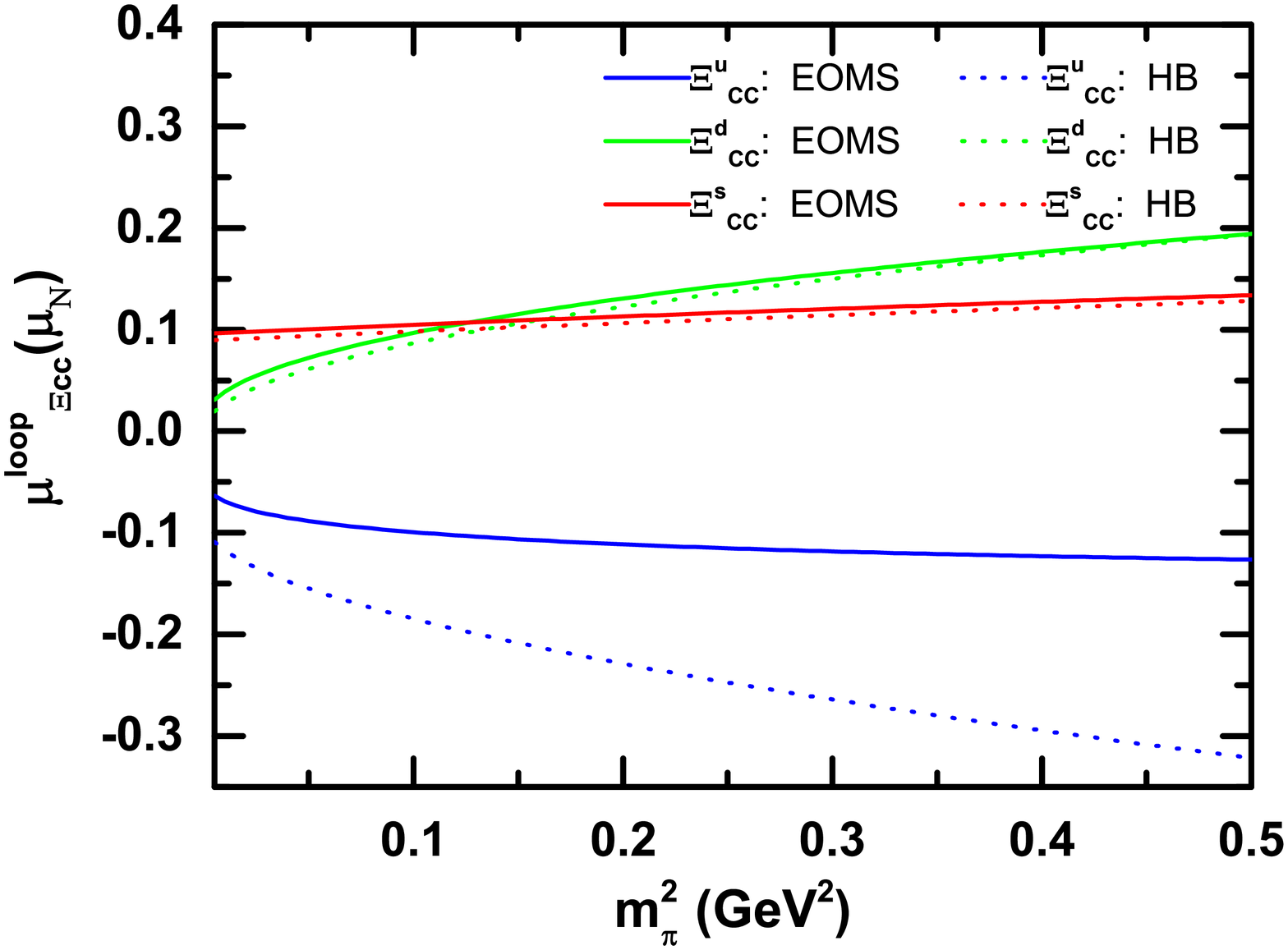}
\end{overpic}
\caption{Loop contributions to the magnetic moments of $\Xi_{cc}$
as a function of $m_{\pi}^2$ for $g_a=0.25$.
}\label{loop1}
\end{figure}

In contrary to the nucleon case~
where the HB and EOMS
results can differ substantially~\cite{Xiao:2018rvd}, for the doubly charmed $\Xi_{cc}$
baryons, the loop contributions are much suppressed. This can
be easily seen from the small $g_a\approx 0.08\sim 0.25$, which is
less than a fifth of  the axial-vector coupling of the nucleon, $g_A=1.26$. As shown in
Fig.~\ref{loop1} the magnetic moments of $\Xi_{cc}^{d}$ and
$\Xi_{cc}^{s}$ receive only small relativistic corrections, while for
$\Xi_{cc}^{u}$ the correction  is slightly larger.  This can serve a
nontrivial test of the ChPT results once more refined lattice QCD data become
available.

One should note that the fits to the lattice QCD simulations are only of exploratory nature.  In the present work, we have not
taken into account finite volume corrections and continuum extrapolations. In addition, because of the limited lattice QCD data, we have
not performed a full study of truncation errors, different from the study of
the magnetic moments of the ground-state octet baryons~\cite{Xiao:2018rvd}.

 \begin{table}[!h]
 \centering
  \caption{Comparison of the magnetic moments of $\Xi_{cc}$ with those predicted by other approaches. Note that the EOMS BChPT results are
  obtained by fitting to the lattice QCD data of Ref.~\cite{Can:2013tna} up to NLO taking $c_1$ and $g_a$ as free LECs. }\label{physicsp}
 \begin{tabular}{l|ccc}
 \hline\hline
$\Psi$  &$\Xi_{cc}^{u}(\mu_{N})$  & $\Xi_{cc}^{d}(\mu_{N})$ &  $\Xi_{cc}^{s}(\mu_{N})$   \\ \hline
QCD sum rule ~\cite{Ozdem:2018uue}          &0.84      &0.46      &0.43                                                \\
HBChPT~\cite{Li:2017vmq}          &$-0.25$     &0.85      &0.78                                               \\
Lattice QCD~\cite{Can:2013tna}           &  -  &0.425      &0.413                                               \\
QM~\cite{Lichtenberg:1976fi}        &$-0.12$    &0.80      &0.69                                               \\
RQM~\cite{JuliaDiaz:2004vh}        &$-0.10$    &0.86      &0.72                                               \\
Skyrmion~\cite{Oh:1991ws}           &$-0.47$      &0.98      &0.59                                                \\
NQM~\cite{Patel:2008xs}          &$-0.20$     &0.79      &0.64                                               \\
$\chi$CQM~\cite{Sharma:2010vv}     &0.006    &0.84      &0.70                                               \\
RTQM~\cite{Faessler:2006ft}        &0.13    &0.72      &0.67                                               \\
NRQM~\cite{SilvestreBrac:1996bg}        &$-0.20$    &0.78      &0.63                                               \\
MIT bag model~\cite{Bose:1980vy}        &0.17    &0.86      &0.84                                               \\
CLP~\cite{Jena:1986xs}        &$-0.154$    &0.778      &0.657                                               \\
EOMS BChPT$^*$           &-      &$0.392(13)$     &$0.397(15) $                                \\
           \hline\hline
 \end{tabular}
 \end{table}

In Table~\ref{physicsp}, we compare the predicted magnetic moments of $\Xi_{cc}$ (Case 1) with those obtained in other approaches. One finds that the theoretical results are very much scattered. Clearly, more investigations are needed to understand the current situation. Such studies may provide vital information on the
nature of these $\Xi_{cc}$ baryons.

A few comments are in order. Clearly, the lattice QCD results of Ref.~\cite{Can:2013tna} and the present BChPT results (based on the same lattice QCD data) are not consistent with
the quark model results. This is somehow surprising because one naively expects that the quark model becomes a better approximation of QCD with increasing quark masses as realized in
lattice QCD simulations. In addition, the rather weak pion mass dependence of the lattice QCD data dictates a $g_a$ much smaller than that predicted by either the quark model or the HADS. This may also be seen as a sign of the inconsistency between the quark model and the lattice QCD simulations. It remains an interesting issue to understand such discrepancies.

\section{SUMMARY}

We calculated the magnetic moments of the $\Xi_{cc}$ baryons in covariant baryon chiral perturbation theory with the extended-on-mass-shell scheme up to the next-to-leading order.
The relevant low-energy constants are determined by fitting to lattice QCD simulations. We showed that the lattice QCD data support an axial-vector coupling $g_a$ smaller than those predicted by
either the quark model or the heavy antiquark diquark symmetry. In addition, we found that relativistic corrections are very small for $\Xi_{cc}^{d}$ and $\Xi_{cc}^{s}$, but relatively
large for $\Xi_{cc}^{u}$. This should be tested by future lattice QCD simulations. On the other hand, we notice that the present lattice QCD results are inconsistent with those of
 the quark model. More  studies, particularly lattice QCD studies,   are therefore in urgent need given the remarkable experimental progress achieved in the last few years.

\section{Acknowledgments}
This work is partly supported by the National Natural Science Foundation of China under Grants No.11522539, 11735003
and the fundamental Research Funds for the Central Universities.

\textit{Note added:} Recently, a study of the electromagnetic form factors of the $\Xi_{cc}$ baryons in the same theoretical framework also appeared in arXiv~\cite{Blin:2018pmj}, focusing more on 
 the $q^2$ dependence of the form factors, rather on their light quark mass dependence. Their predicted magnetic moments, with a $|g_a|=0.2$, are consistent with ours within uncertainties.

\end{document}